\documentclass[twocolumn, showpacs,preprintnumbers,amsmath,amssymb]{revtex4}

\usepackage{graphicx}

\begin{document}


\title{Decoherence of intermolecular entanglement in exchange-coupled nanomagnets}

\author{A. Szallas$^{1,2}$ and F. Troiani$^{1}$}

\affiliation{$^1$S3, Istituto Nanoscienze-CNR, Via G. Campi 213A, I-41125 Modena\\
$^2$Dipartimento di Fisica, Universit\`a di Modena e Reggio Emilia, via G. Campi 213/a, 41100 Modena, Italy}

\date{\today}

\begin{abstract}

We theoretically investigate the hyperfine-induced 
decoherence in a pair of spin-cluster qubits, consisting of two exchange-coupled 
heterometallic wheels. 
We identify two distinct regimes in the decoherence of intermolecular 
entanglement and show that this can be substantially recovered through 
dynamical decoupling.
Different chemical elements and physical processes are responsible 
for the decoherence of the singlet-triplet superposition, resulting in 
a wider tunability of its decoherence time.

\end{abstract}

\pacs{75.50.Xx, 03.65.Yz, 03.67.Bg}


\maketitle

\begin{figure}[ptb]
\begin{center}
\includegraphics[width=7.5cm]{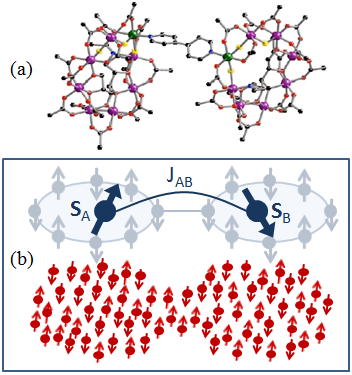}
\end{center}
\caption{(color online) (a) Chemical structure of the (Cr$_7$Ni)$_2$ 
dimer \cite{PhysRevLett.104.037203}, 
as determined by X-ray diffraction: Cr (purple), Ni (green), F (yellow), 
O (orange), C (black), N (cyan); the H nuclei have been omitted for the 
sake of clarity. 
(b) Schematics of the spin system and environment. 
The state of the two electron-spin rings (faint blue), is mapped onto 
that of two exchange-coupled 1/2 spins, $ S_A $ and $ S_B $ (blue). The 
electron spins interact with a bath of $ N_n = 312 $ nuclear spins (red).}
\label{fig1}
\end{figure}
Molecular nanomagnets (MNs) represent a rich class of spin clusters, whose 
properties can be widely tailored by chemical synthesis \cite{book_MN}.
Weak interactions between well defined molecular units have also been 
demonstrated \cite{WW02} and can be made tunable by introducing different 
intermolecular linkers \cite{NatureNanotech.4.173}.
This has recently allowed to demonstrate equilibrium-state entanglement
between pairs of heterometallic wheels \cite{PhysRevLett.104.037203}.
Quantum mechanical features, such as free \cite{PhysRevLett.98.057201,
PhysRevLett.102.087603} and forced \cite{bertaina08,PhysRevLett.102.050501,
PhysRevLett.101.147203} oscillations, have also emerged in the coherent 
dynamics of MNs. 
The tunable coupling between nanomagnets, combined with the coherent 
manipulation of their spin state by pulsed EPR, will possibly enable the 
experimental investigation and the controlled generation of intermolecular 
coherence and entanglement. 
Besides being of fundamental interest, these issues are pivotal to the 
implementation of quantum information processing with MNs \cite{leuenberger,
PhysRevLett.90.047901,PhysRevLett.94.190501,PhysRevLett.101.217201}.

The main sources of electron spin decoherence in MNs are represented by
intermolecular dipole-dipole interactions \cite{PhysRevLett.97.207206},
spin-phonon \cite{phonons} and hyperfine coupling \cite{prokofiev,coish}. 
While the former two contributions can be limited by 
dilution of the molecular ensemble and low temperatures 
\cite{PhysRevLett.98.057201}, the suppression of hyperfine-induced 
decoherence might require more sophisticated dynamical decoupling 
techniques \cite{PhysRevA.58.2733}.
In spite of its fundamental and applicative interest, still
little is known on intermolecular coherence in coupled nanomagnets
and on its robustness with respect to hyperfine interactions.
Here we address this problem in the case of two spin-cluster 
qubits, each consisting of a Cr$_7$Ni heterometallic wheel 
(Fig. \ref{fig1}(a)) \cite{nanotech}.
Being only the ground-state doublet of each wheel 
involved in the low-temperature dynamics, the electron spin state of 
the (Cr$_7$Ni)$_2$ dimer can be mapped onto that of two exchange-coupled 
spins $ S_A = S_B = 1/2 $ (Fig. \ref{fig1}(b)). 
Essentially, decoherence is due to the dependence of the slow nuclear 
bath dynamics on the state of the electron spins, and to 
the resulting correlations between electron and nuclear spins. 
The features that characterize the decoherence of different linear 
superpositions can however remarkably differ from one another \cite{RevModPhys.75.715}. 
We indeed find similar differences between the decoherence of two 
prototypical linear superpositions $ ( |\Phi_1\rangle + 
|\Phi_2\rangle ) / \sqrt{2} $, with $ |\Phi_i\rangle $ the electron-spin 
eigenstates of the dimer.
The first linear superposition, with components 
$ |\Phi_1\rangle = |\Uparrow_A,\Uparrow_B\rangle $ and $ |\Phi_1\rangle 
= |\Downarrow_A , \Downarrow_B\rangle $, corresponds to an entangled 
state of the two Cr$_7$Ni molecules. 
Here, electron-nuclear correlations are shown to arise mainly from the 
opposite chemical shifts of the H nuclei induced by $ |\Phi_1\rangle $ and 
$ |\Phi_2\rangle $.
In the second case, the components are instead $ |\Phi_{1,2}\rangle = 
( |\Uparrow_A,\Downarrow_B\rangle \pm |\Downarrow_A,\Uparrow_B\rangle ) 
/ \sqrt{2}$.
Both the singlet and triplet are characterized by a vanishing expectation 
value of the electron spin projection, and therefore induce no chemical 
shift of the nuclei. 
We find that, as a consequence, the electron-nuclear correlations result from 
processes that 
are second-order in the hyperfine interaction (i.e. electron-spin mediated 
transitions between pairs of nuclei) and selectively involve the F nuclei, 
localized close to the electron spins. 

\begin{figure}[ptb]
\begin{center}
\includegraphics[width=8.5cm]{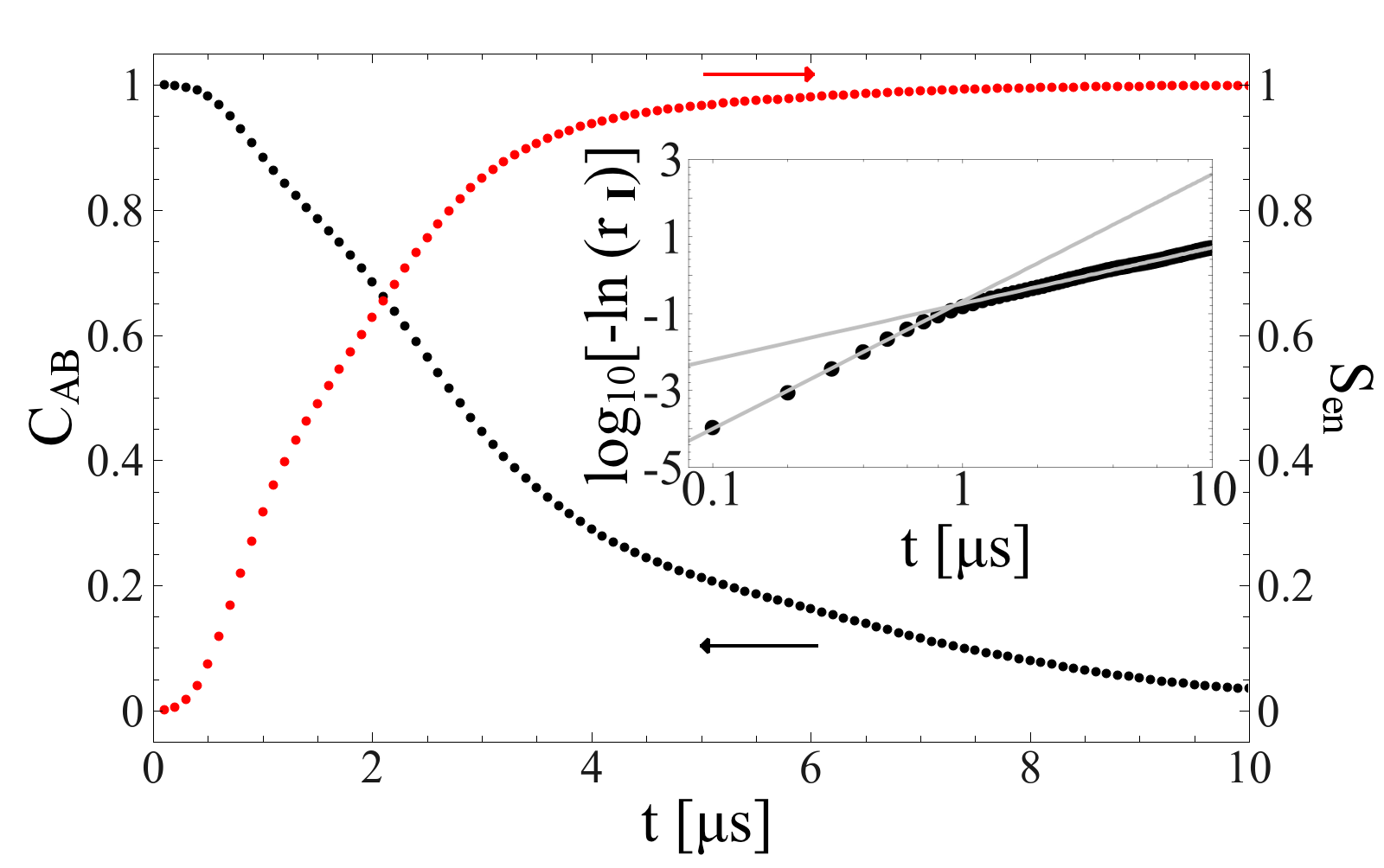}
\end{center}
\caption{(color online)
Simulated time evolution of the entanglement between $ {\bf S}_A $ and 
$ {\bf S}_B $, as quantified by the concurrence, $ \mathcal{C}_{AB} 
( \rho_e^{BS} ) $ (black circles, left axis). 
The entanglement between the electron and nuclear spins, for a randomly
generated state of the nuclear bath $ | \bf I \rangle $, is quantified by 
the von Neumann entropy (red circles, right axis). 
Figure inset: calculated time-dependence of $ r_{\bf I} $ (black circles)
and fitting curves $ f(t) = \exp\{-(t/\tau_d)^\alpha\} $ (gray lines), 
with $ \alpha = 3.31 $ and $ \tau_d = 0.70\,\mu $s and $ \alpha = 
1.46 $ and $ \tau_d = 0.75\,\mu $s.}
\label{fig2}
\end{figure}
{\it The model --- } In the (Cr$_7$Ni)$_2$ dimer, the spin system and 
environment consist of $ N_e = 16 $ electron spins and $ N_n = 312 $ nuclear 
spins, respectively (Fig. \ref{fig1}(b)).
Each Cr$_7$Ni wheel is modeled by a 
ring of eight electron spins -- one for each of the magnetic ions -- with $ s_{1-7} =
s_{\rm Cr} = 3/2 $ and $ s_8 = s_{\rm Ni} = 1 $. 
The two lowest eigenstates of each single-ring Hamiltonian form a Kramers 
doublet: 
$ | \!\Uparrow_\chi\!   \rangle \!\equiv\! | S_\chi\! =\! 1/2, M_\chi\! =\!  1/2 \rangle $ 
and 
$ | \!\Downarrow_\chi\! \rangle \!\equiv\! | S_\chi\! =\! 1/2, M_\chi\! =\! -1/2 \rangle $,
where $ \chi = A , B $.
At zero magnetic field this doublet is separated from the first 
excited quadruplet by an energy gap of $ \Delta \simeq 20\,$K 
\cite{PhysRevLett.104.037203}.
At temperatures $ T \ll \Delta $, the state and the couplings of each ring 
can thus be mapped onto those of an effective $ 1/2 $ spin.
This also applies to the intermolecular exchange induced by the bipyridine 
linker in the (Cr$_7$Ni)$_2$ dimer (Fig. \ref{fig1}(b)) \cite{PhysRevLett.104.037203}, 
and to the hyperfine
couplings to the nuclear spins \cite{PhysRevB.77.054428}.
In the presence of an applied magnetic field the effective two-spin Hamiltonian
reads:
\begin{equation}\label{hab}
\mathcal{H}_{AB} = J_{AB} {\bf S}_A \cdot {\bf S}_B + g \mu_B \sum_{\chi = A,B} 
{\bf S}_\chi \cdot {\bf B} ,
\end{equation}
with $ J_{AB} \ll \Delta $.
The spin bath consists of $ N_n = 312 $ nuclear spins (6 F and 306 H, with 
$ I_{\rm F} = I_{\rm H} = 1/2 $), whose positions are derived from X-ray 
crystallography. The nuclear spins $ {\bf I}_k $ interact with each other and 
with the electron spins via dipole-dipole interactions, which are accounted for 
within the point-dipole approximation. 

Energy relaxation of the electron spin state, arising from transitions between 
different eigenstates $ | S , M \rangle $ (with $ {\bf S} = {\bf S}_A + {\bf S}_B $), 
is strongly inhibited, being the differences 
between the eigenvalues of $ \mathcal{H}_{AB} $ are much larger than 
both the excitation energies of the nuclear spins and of the hyperfine couplings.
In such pure-dephasing regime, the electron-spin decoherence 
is induced by the dependence of the nuclear-spin dynamics on the electron spin 
state, and by the resulting electron-nuclear correlations.
In particular, an initial state 
\begin{equation}\label{eqis}
 |\Psi_0\rangle\! =\! |\Psi_{e0}\rangle\! \otimes\! |\Psi_{n0}\rangle\! =\! \frac{1}{\sqrt{2}}
   \left( | S_1 , M_1 \rangle\! +\! | S_2 , M_2 \rangle \right) 
     \!\otimes\! | {\bf I} \rangle 
\end{equation}
(where $ |\Psi_{e0}\rangle $ and $ |\Psi_{n0}\rangle = | {\bf I} \rangle = 
\otimes_{k=1}^{N_n} | I_k^z \rangle $ are the electron and nuclear components, 
respectively) evolves into 
\begin{equation} \label{leq02}
 |\Psi_t \rangle\!\! =\!\! \frac{1}{\sqrt{2}}\!
    \left[ | S_1 , M_1 \rangle\!\! \otimes\!\! 
| {\bf I}_1 (t) \rangle \! +\! e^{i\phi_{\bf I} (t)} 
      | S_2 , M_2 \rangle\!\! \otimes\!\! 
| {\bf I}_2 (t) \rangle \right]\!\! .
\end{equation}
Here, $ | {\bf I}_1 (t) \rangle $ and
$ | {\bf I}_2 (t) \rangle $ are the states of
the nuclear bath conditioned upon the electron spins being in the 
$ | S_1 , M_1 \rangle $ or
$ | S_2 , M_2 \rangle $ states,
respectively.
The reduced density matrix of the electron spins is thus given by:
\begin{eqnarray} \label{rhoe}
\rho_e (t) 
& = & {\rm Tr}_n \left\{ |\Psi_t \rangle\langle \Psi_t | \right\} = \frac{1}{2} \sum_{ k = 1 , 2 }  
| S_k , M_k \rangle\langle S_k , M_k | \nonumber\\
 & + &
\frac{r_{\bf I} (t)}{2} \left[ 
| S_2 , M_2 \rangle\langle S_1 , M_1 | 
e^{i\phi_{\bf I} (t)} + {\rm h. c.}
\right] ,
\end{eqnarray}
where $ r_{\bf I} = | \langle {\bf I}_1 | {\bf I}_2 \rangle | $ and 
$ \phi_{\bf I} = \arg\{ \langle {\bf I}_1 | {\bf I}_2 \rangle \} $ result from the 
dynamics of the nuclear bath.

This is derived from the effective Hamiltonian \cite{PhysRevB.74.195301,PhysRevB.77.054428}: 
\begin{equation} \label{leq01}
\mathcal{H}_{\rm eff} = \sigma_z^e \otimes \mathcal{H}_n^{(e)} + 
\mathcal{I}^e \otimes \mathcal{H}_n^{(i)} ,
\end{equation}
being $ | \Psi_t \rangle = \exp\{ - i \mathcal{H}_{\rm eff} 
t / \hbar \} | \Psi_0 \rangle $.
Here, the electron and nuclear-spin operators, are given by:
$ \sigma_z^e \equiv 
| S_1 , M_1 \rangle\langle S_1 , M_1 | -
| S_2 , M_2 \rangle\langle S_2 , M_2 | $,
$ {\mathcal I}^e \equiv 
| S_1 , M_1 \rangle\langle S_1 , M_1 | +
| S_2 , M_2 \rangle\langle S_2 , M_2 | $,
and 
\begin{equation} \label{eqeffham}
\mathcal{H}_n^{(\alpha)} = \sum_{k=1}^{N_n} A^{(\alpha)}_k I^z_k + 
\sum_{k,l=1}^{N_n} \left[ B^{(\alpha)}_{kl} I^z_k I^z_l + C^{(\alpha)}_{kl} I^+_k I^-_l \right] ,
\end{equation}
with $\alpha = e,i$.
The intrinsic nuclear Hamiltonian $ \mathcal{H}_n^{(i)} $ includes the 
nuclear Zeeman and the secular (i.e. Zeeman-energy conserving) terms of the 
dipole-dipole interactions between nuclei.
The extrinsic Hamiltonian $ \mathcal{H}_n^{(e)} $ accounts for the Overhauser
effect (couplings $A^{(e)}_k$) and for the interactions between nuclei mediated 
by virtual excitations of the electron-spins ($B^{(e)}_k$ and $C^{(e)}_k$).

{\it Bell-state decoherence ---}
In order to investigate the decoherence of intermolecular entanglement,
we consider the case where the components of $ | \Psi_{e0} \rangle $ 
(Eq. \ref{eqis}) are $ S_1 = M_1 = 1 $ and $ S_2 = - M_2 = 1 $,
which corresponds to the spins ${\bf S}_A$ and ${\bf S}_B$ being in the 
Bell state 
$
 | \Psi_{e0} \rangle  = \frac{1}{\sqrt{2}}
                \left( | \! \Uparrow_A   , \Uparrow_B   \rangle + 
                       | \! \Downarrow_A , \Downarrow_B \rangle \right).
$
The simulated build up of correlation between the states of electron and 
nuclear spins and the resulting loss of entanglement between ${\bf S}_A$ 
and ${\bf S}_B$ is reported in Fig. \ref{fig2}.
The entanglement between two qubits is quantified by the concurrence 
\cite{RevModPhys.80.517}, which, for the density matrix $ \rho_e $ 
(Eq. \ref{rhoe}), reduces to:
\begin{equation} \label{eqent1}
\mathcal{C}_{AB} ( \rho_e ) = r_{\bf I} .
\end{equation} 
The von Neumann entropy \cite{RevModPhys.80.517} can be used to quantify 
entanglement between electron and nuclear spins. Its expression, for 
$ | \Psi_t \rangle $ (Eq. \ref{leq02}), turns out to be:
\begin{equation} \label{eqent2}
S_{en} ( \rho_e ) = - \frac{1}{2} \sum_{\mu = \pm 1} (1 +\mu r_{\bf I}) 
\log_2 \left[ \frac{1}{2} (1 +\mu r_{\bf I})\right] .
\end{equation} 
As $ \rho_e $ evolves from the initial linear superposition to a 
statistical mixture of $ | \! \Uparrow_A, \Uparrow_B \rangle $ and 
$ | \! \Downarrow_A , \Downarrow_B \rangle $, $ \mathcal{C}_{AB} $ (black
circles) and  $ S_{en} $ (red circles) tend to 0 and 1, respectively.
This clearly shows the mutually exclusive nature of electron-nuclear and 
intermolecular entanglement. 
Besides, two distinct regimes can be clearly identified in the entanglement 
decoherence: for times $ t \lesssim 1 \,\mu$s, $ r_{\bf I} (t) $ follows
an exponential decay $ f(t) = \exp\{-(t/\tau_d)^\alpha\} $, with $ \alpha 
= 3.31 $ and $ \tau_d = 0.70\,\mu $s; for $ t \gtrsim 1 \,\mu$s, instead, 
$ \alpha = 1.46 $ and $ \tau_d = 0.75\,\mu $s (figure inset).
\begin{figure}[ptb]
\begin{center}
\includegraphics[width=8.5cm]{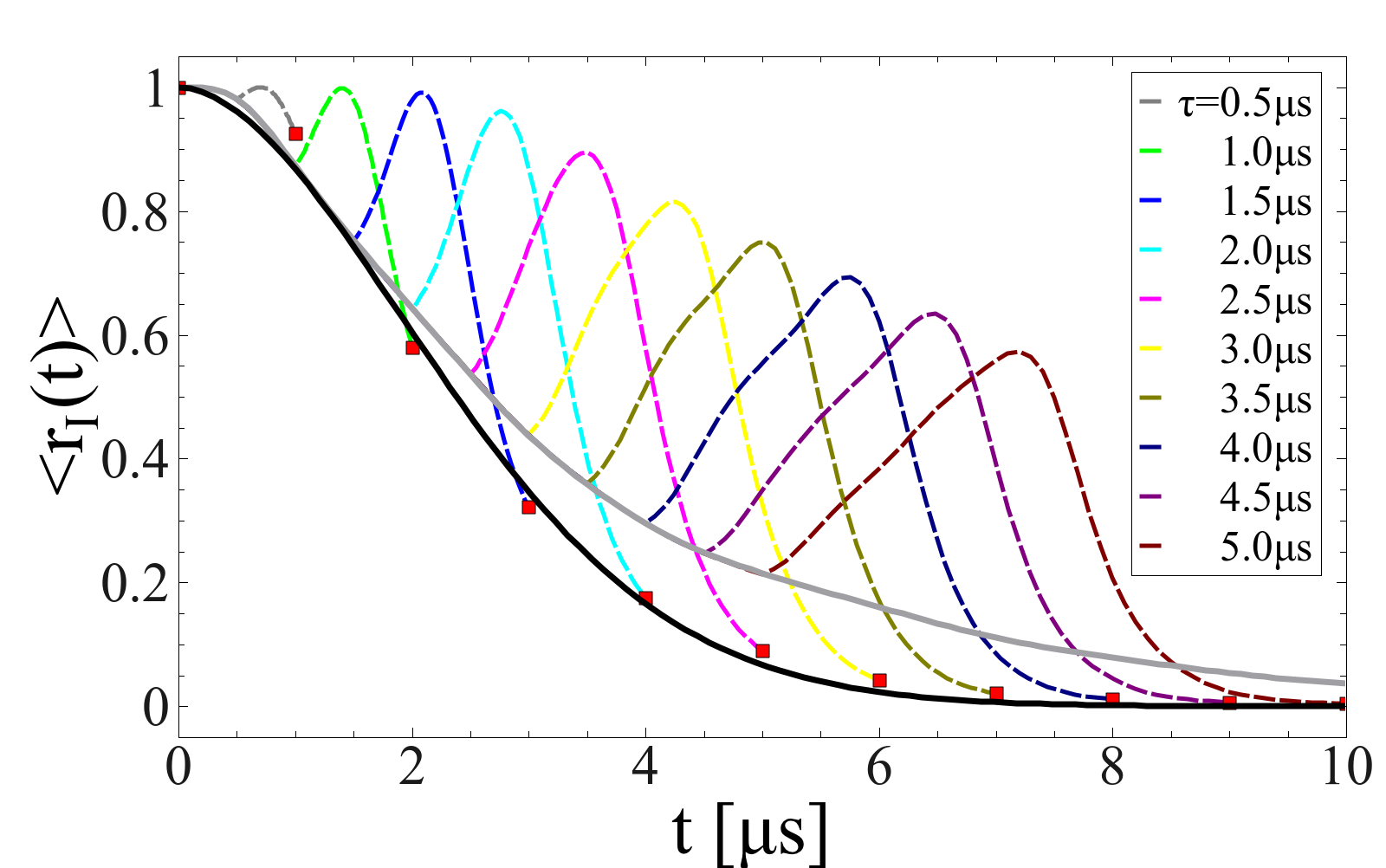}
\end{center}
\caption{(color online) Time evolution of $ r_{\bf I} $, averaged over 
$ N = 100 $ randomly generated initial states $ | {\bf I} \rangle $, 
in the absence or in the presence of a $ \pi $-pulse applied to the 
electron spins at $ t=\tau $ (solid-gray and dotted curves, respectively;
different colors correspond to different values of $ \tau $).
The values $ \langle r_{\bf I} (t=2\tau) \rangle $ for different $ \tau $
(red squares) are fitted by $ f(t) = \exp\{-(t/\tau_d)^\alpha\} $,
with $ \alpha = 1.8 $ and $ \tau_d = 2.9\,\mu $s.}
\label{fig3}
\end{figure}

For temperatures much larger than the nuclear Zeeman splitting, the density
matrix of the nuclear bath resembles the maximally disordered mixture of 
states $ | {\bf I} \rangle $: $ \rho_{n0} = \sum_{\bf I} | {\bf I} \rangle
\langle {\bf I} | / 2^{N_n}$.
While $ r_{\bf I} (t) $ is largely independent on the initial state 
$ | {\bf I} \rangle $ of the nuclear bath, this is not the case for the 
phase $ \phi_{\bf I} $, which depends on the Overhauser field induced by 
the nuclei (see Eq. \ref{eqeffham}): $ B_{\bf I}^n = - \sum_{k=1}^{N_n} 
A^{(e)}_k I_k^z / ( \mu_B \bar{g}_z ) $ ($ \bar{g}_z $ is the effective 
$g$ factor of each Cr$_7$Ni ring).
The static component of the Overhauser field and the resulting 
inhomogeneous broadening produce a fast dephasing, that can however be 
eliminated by spin-echo techniques \cite{PhysRevLett.98.057201}. 
The slow dynamics of the nuclear bath however also results in a fluctuating 
component of the field, with a potentially irreversible contribution to the 
phase $ \phi_{\bf I} (t) $ and to the electron-spin decoherence. 
The simulated dynamics however shows that such contribution affects only 
marginally entanglement decoherence in the (Cr$_7$Ni)$_2$ dimer:
\begin{equation}
\mathcal{C}_{AB} ( \rho_e^{BS} ) 
= \langle \, r_{\bf I} \, e^{i\phi_{\bf I}} \rangle 
\simeq \langle \, r_{\bf I} \rangle ,
\end{equation}
where 
$ 0 \le | \langle r_{\bf I} e^{i\phi_{\bf I}} \rangle | - \langle r_{\bf I} 
\rangle \le 0.012 $ for $ t \le 10\,\mu$s, and the average is performed on 
a set of randomly generated initial states $ | {\bf I} \rangle $.

In order to gain further insight into the decoherence process and underlying
dynamics, we isolate the role played by the specific chemical elements and 
the different physical mechanisms. To this aim, we consider the difference 
between $ | \langle r_{\bf I} e^{i\phi_{\bf I}} \rangle | $ resulting from 
the full dynamics and the same quantity obtained by neglecting part of the 
terms in $ \mathcal{H}_{\rm eff} $. 
As detailed in Table \ref{table}, the contribution of the H (F) nuclei to the 
decoherence of intermolecular entanglement is predominant (negligible).
Amongst the hyperfine coupling mechanisms, the effect of the chemical shift 
(i.e. the renormalizations of the nuclear Zeeman energies induced by the 
electron spins) largely prevails over the electron-spin mediated interactions 
between the nuclei. 
The relative importance of these contributions changes drastically in the 
case of the singlet-triplet decoherence (see below).

The correlations between electron and nuclear spins described above can be 
partially reversed by dynamical decoupling schemes \cite{PhysRevA.58.2733}. 
In Fig. \ref{fig3} we show for example the effect on entanglement decoherence 
of a flip of the electron spins $ {\bf S}_{\chi = A,B} $ at a time $t=\tau$, 
that can be induced by an EPR $\pi$-pulse.
This ideally transforms the two components of $ | \Psi_{e0} \rangle $ into 
one another ($ | \Uparrow_A , \Uparrow_B \rangle \longleftrightarrow 
| \Downarrow_A , \Downarrow_B \rangle $), and thus $ \sigma_z^e 
\longrightarrow - \sigma_z^e $ in $ \mathcal{H}_{\rm eff} $ at $ t = \tau $. 
Such inversion of the extrinsic part of the Hamiltonian tends to undo the 
electron-nuclear correlations; $ \langle r_{\bf I} \rangle $  
correspondingly increase for $ t > \tau $, with a maximum localized around 
$ t \simeq \sqrt{2} \tau $ (colored lines) \cite{PhysRevLett.98.077602}. 
The above mentioned $\pi$-pulse is used within the Hahn-echo sequence, in 
order to cancel the effect of inhomogeneous broadening in free-induction decay 
experiments and to induce an observable spin echo at time $ t = 2\tau $. 
At that time, electron-nuclear correlations have built up again (red squares),
determining values of $ \langle r_{\bf I} \rangle $ which are lower than those 
obtained in the absence of the $ \pi-$pulse (brown curve). More complex 
dynamical decoupling schemes are thus required in order to achieve simultaneously 
the removal of inhomogeneous broadening and electron-nuclear entanglement
and to make the latter effect experimentally observable \cite{dyde}.

\begin{figure}[ptb]
\begin{center}
\includegraphics[width=8.5cm]{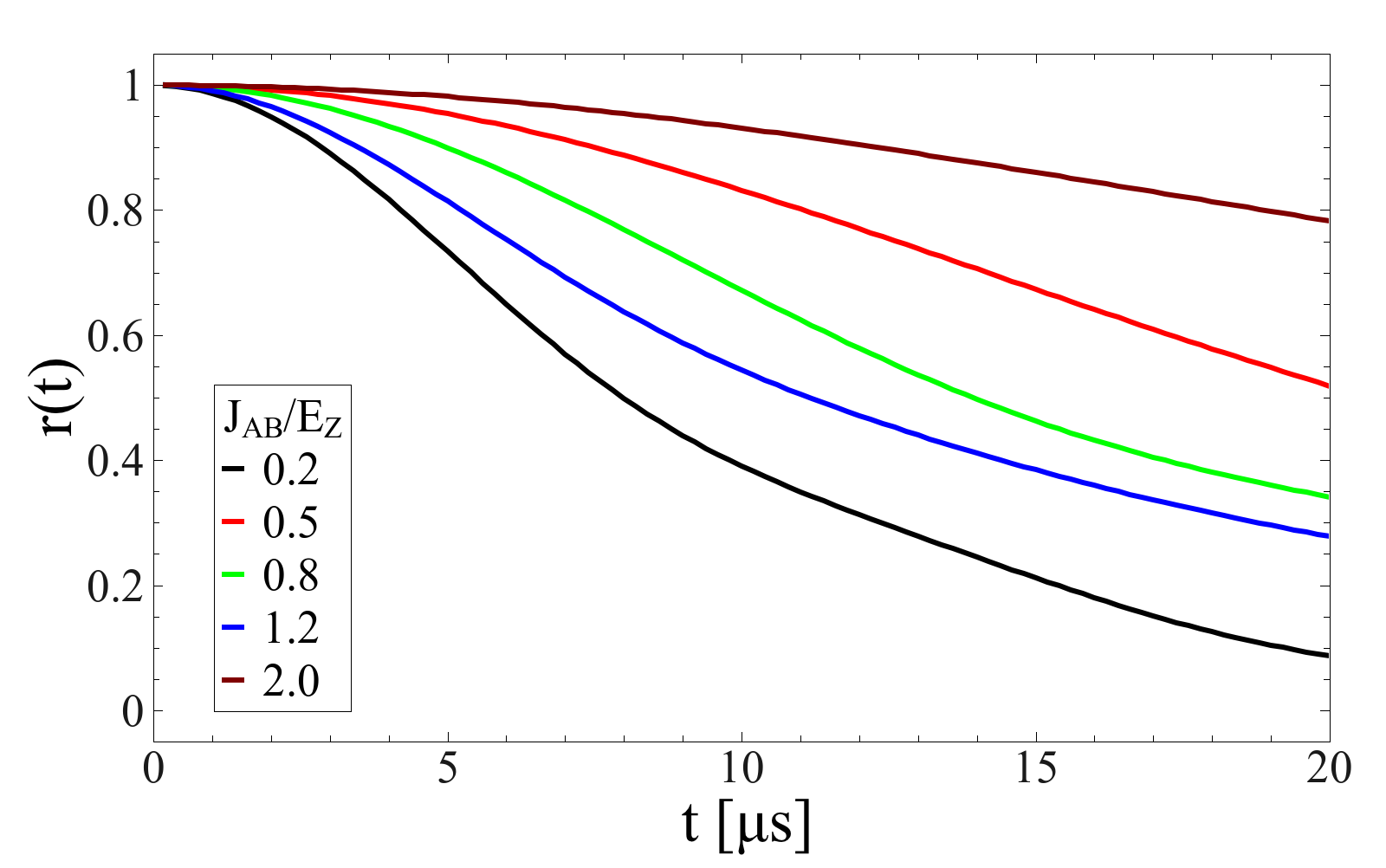}
\end{center}
\caption{(color online) Singlet-triplet decoherence as a function of the 
intermolecular exchange $J_{AB}$ the Zeeman energy $ E_Z = g \mu_B B_z $
(Eq. \ref{hab}). 
The function $ r = \langle r_{\bf I} e^{i \phi_{\bf I}} \rangle $ has been 
averaged on $ N = 128 $ randomly generated initial states 
$ | {\bf I} \rangle $.}
\label{fig4}
\end{figure}
{\it Singlet-triplet decoherence ---} 
In order to complement the results on the entangled state decoherence,
we consider the case of a fundamentally different linear combination, 
namely 
that between the singlet and triplet states ($ S_1 = 0 $, $ S_2 = 1 $ and $ M_1 = M_2 = 0 $ 
in Eq. \ref{eqis}).
The reduced density matrix, derived from Eq. \ref{leq02}, now reads:
$
\rho_{e} (t) = \frac{1}{2} [ (1+r) 
| \Uparrow_A , \Downarrow_B \rangle\langle \Uparrow_A , \Downarrow_B | +
(1-r) 
   | \Downarrow_A , \Uparrow_B \rangle\langle \Downarrow_A , \Uparrow_B | ] 
$
(with $ r \equiv \langle r_{\bf I} e^{i\phi_{\bf I}} \rangle $),
and is always factorizable into the states of $ {\bf S}_A $ and $ {\bf S}_B $: 
$\mathcal{C}_{AB} ( \rho_e ) = 0$. 

\begin{table}
\begin{tabular}{|c|cccc|}
\hline
\ \ \ \ & \ \ Chemical \ \ & \ \ Terms \ \ & \ \ Terms \ \ & \ \ $ \ \ \delta r \ \ $ \\
       & elements & in $ \mathcal{H}^{(i)}_n $ & in $ \mathcal{H}^{(e)}_n $ & \\
\hline
BS      & H & $ A^{(i)}_k $, $ B^{(i)}_{kl} $, $ C^{(i)}_{kl} $ & $ A^{(e)}_k $ & 0.02 \\
ST & F & $ A^{(i)}_k $ & $ A^{(e)}_k $, $ B^{(e)}_{kl} $, $ C^{(e)}_{kl} $ & 0.04 \\
\hline
\end{tabular}
\caption{\label{table} Main contributions to the decoherence of the Bell state
(BS) and of the singlet-triplet superposition (ST). 
The key quantity is
$ \delta r \equiv \max_t | \langle r_{\bf I}^* \rangle - 
\langle r_{\bf I} \rangle | $, where $ r_{\bf I}^* $ is computed by 
keeping in $ \mathcal{H}_{\rm eff} $ (Eq. \ref{leq01}) only the 
chemical elements and the terms specified in the central columns; 
the considered 
time interval is $ 0 \le t \le 10\,\mu$s.}
\end{table}
The features of the singlet-triplet decoherence are fundamentally different 
from those emerged in the case of the Bell state. 
As shown in Fig. \ref{fig4}, the time-evolution of $ r $ is strongly 
dependent on the intermolecular exchange $J_{AB}$ (Eq. \ref{hab}).
In particular, the decoherence rate is on average much smaller than that of the 
Bell state, and decreases for increasing energy difference between the singlet 
state and each of the triplet states.
An analogous dependence (not shown here) is found with respect to the 
magnetic field. 
These features are fully consistent with the identification of transitions 
between F nuclei, mediated by virtual transitions of the electron-spin state, 
as the dominant contribution to the overall decoherence (see Table \ref{table}).
The dynamics of the H nuclei, though present, is largely independent on the 
electron-spin state, and therefore doesn't result in the built up of the 
electron nuclear correlations.
As anticipated in the introduction, such correlations arise in the case of the 
entangled state from the fact that the two components ($| \Uparrow_A , \Uparrow_B 
\rangle$ and $| \Downarrow_A , \Downarrow_B \rangle$) give rise to opposite 
chemical shifts. Instead, the chemical shift induced by the singlet and $M=0$ 
triplet states vanishes, being $ \langle S , 0 | s_{i,z}^\chi | S , 0 \rangle 
= 0 $. The dependence of the nuclear spin dynamics on the electron state 
therefore relies essentially on second-order processes and on the contribution 
of the F nuclei, that are localized closer to the electron spins of the dimer 
(Fig. \ref{fig1}(a)).

{\it Conclusions ---}
While a number of additional aspects need to be considered in view of their
possible relevance in experiments (for example, those related to the hyperfine 
interaction with the nuclei in the solvent, or to the intialization and 
manipulation of the electron spin state of the dimer), some fundamental features 
have emerged from the presented simulations. In particular, we have identified the 
clear occurrence of two regimes in the build up of electron-nuclear correlations,
and of the resulting decoherence of intermolecular entangled. 
The comparison with the case of singlet-triplet decoherence shows that the 
decoherence in the low-energy subspace of the dimer can be induced by distinct 
chemical elements and physical processes, depending on the components of the 
initial linear combination. Finally, as demonstrated in other spin systems, 
dynamical decoupling might allow to disentangle electron and nuclear spins, thus 
leading to a substantial recovery of the intermolecular coherence. 

We are grateful to M. Affronte and V. Bellini for fruitful discussions; we 
thank M. Rontani and D. Prezzi for a careful reading of the manuscript. 
We acknowledge financial support from the Italian CNR-INFM under SEED 2008 
and the EU under MolSpinQIP. 


\end{document}